\documentclass[12pt,english,english,prd,nofootinbib]{revtex4-1}
\usepackage[T1]{fontenc}
\usepackage[latin9]{inputenc}
\usepackage{float}
\usepackage{amsmath}
\usepackage{amsthm}
\usepackage{graphicx}
\usepackage{esint}

\makeatletter

\providecommand{\tabularnewline}{\\}

\numberwithin{equation}{section}
\numberwithin{figure}{section}

\usepackage{amsfonts}

\makeatother

\usepackage{babel}
\begin{document}

\title{Holographic Wilson loops, Hamilton-Jacobi equation and regularizations}

\author{Diego Pontello}

\email{diego.pontello@ib.edu.ar }

\author{Roberto Trinchero$^{a}$}

\email{trincher@cab.cnea.gov.ar}

\address{Instituto Balseiro, Centro Atómico Bariloche, 8400 San Carlos de
Bariloche, Argentina.}

\affiliation{$^{a}$CONICET, Rivadavia 1917, 1033 Cap. Fed., Argentina.}
\begin{abstract}
The minimal area for surfaces whose border are rectangular and circular
loops are calculated using the Hamilton-Jacobi (HJ) equation. This
amounts to solve the HJ equation for the value of the minimal area,
without calculating the shape of the corresponding surface. This is
done for bulk geometries that are asymptotically AdS. For the rectangular
countour, the HJ equation, which is separable, can be solved exactly.
For the circular countour an expansion in powers of the radius is
implemented. The HJ approach naturally leads to a regularization which
consists in locating the countour away from the border. The results
are compared with other regularization which leaves the countour at
the border and calculates the area of the corresponding minimal surface
up to a diameter smaller than the one of the countour at the border.
The results do not coincide, this is traced back to the fact that
in the former case the area of a minimal surface is calculated and
in the second the computed area corresponds to a fraction of a different
minimal surface whose countour lies at the boundary.
\end{abstract}
\maketitle

\section{Introduction}

The relation between large $N$ gauge theories and string theory \cite{'tHooft:1973jz}
together with the AdS/CFT correspondence \cite{Maldacena:1997re,Witten:1998qj,Gubser:1998bc,malda:1999ti}
have opened new insights into strongly interacting gauge theories.
The application of these ideas to QCD has received significant attention
since those breakthroughs. From the phenomenological point of view,
the so called AdS/QCD approach has produced very interesting results
in spite of the strong assumptions involved in its formulation \cite{Gubser:1999pk,Sakai:2004cn,Da_Rold:2005zs,Erlich:2005qh,Polchinski:2000uf,Csaki:2006ji}.
It seems important to further proceed investigating these ideas and
refining the current understanding of a possible QCD gravity dual.

In the holographic approach, the vacuum expectation value of the Wilson
loop is obtained by minimizing the Nambu-Goto (NG) action for a loop
lying in the boundary space \cite{Maldacena:1998im,Rey:1998ik}. This
is known to work in the strictly AdS case, i.e. for a conformal boundary
field theory. In this work it is assumed that this procedure also
works in the non-conformal-QCD case provided an adequate 5-dimensional
background metric is chosen.

In this work the minimal area is computed by solving the Hamilton-Jacobi
equation. This approach has the advantage that the minimal area can
be obtained without solving the equations of motion. It amounts to
study the variation of the minimal area under changes in the location
and shape of the countour. This approach naturally leads to a regularization
which consists in moving the countour into the bulk out of the border.
This HJ-regularization was also considered in \cite{Drukker:1999zq}
for the AdS case. In that reference another regularization was also
employed, which consists in locating the countour at the border but
computing the area only up to a diameter smaller than that of the
countour. This approach will be referred to as $\epsilon$-scheme.
It was shown that the result for smooth surfaces computed using both
schemes coincide except in what respects to zig-zag symmetry. The
HJ-scheme respects this symmetry but the $\epsilon$-scheme does not.
In the present work, it is shown that for the non-AdS case the results
for the coefficients of the expansion in powers of the diameter of
the circular countour of the NG action do not coincide for both schemes,
even for regular surfaces. The origin of this difference between both
approaches is that, in the HJ-scheme, boundary conditions for the
minimal surface are taken at its border, i.e. where the base of the
loop lies. In the $\epsilon$-scheme boundary conditions are taken
at the space border, which is not the location of the calculated area
border.

The features and results of this work are summarized as follows:
\begin{itemize}
\item The HJ approach is employed for the calculation of minimal areas of
rectangular and circular loops in asymptotically AdS spaces.
\item For the case of the rectangular loop the HJ equation is separable
and can be solved exactly.
\item For the case of the circular loop an expansion of the Nambu-Goto (NG)
on-shell action in powers of the radius of the loop is implemented.
At each order the relevant differential equation is linear and solvable
up to the calculation of an integral.
\item The HJ approach naturally leads to a regularization that consists
in locating the loop countour away from the border. The substraction
is implemented following \cite{Maldacena:1998im} as extended to the
non-AdS case in \cite{Quevedo:2013iya}.
\item The two regularizations considered in \cite{Drukker:1999zq} are applied
in this case. One of them is the one mentioned above that fits naturally
in the HJ approach. The other one considers a minimal surface whose
countour is at the border and computes the area of the surface up
to a diameter smaller than the one of the countour at the border.
\item The results for the expansion coefficients of the NG on-shell action
in powers of the radius are considered. They do not coincide for the
two regularizations mentioned above. This discrepancy is investigated
in detail and has its origin in the divergence of the metric coefficients
at the border.
\end{itemize}
This paper is organized as follows. Section 2 defines the bulk metrics
to be considered and recalls the NG action. Section 3 deals with the
rectangular loop in the HJ approach. Section 4 studies the circular
loop in the HJ approach and the approximate solution of the HJ equation
as a power series in the loop's radius. Section 5 deals with the substraction
scheme and its explicit computation. Section 6 compares both regularizations
and explains the origin of the discrepancy between both. Sections
7 presents some concluding remarks. In addition two appendices are
included, one of them giving explict expressions of the expansion
coefficients mentioned above and the other showing the source of the
differences between both regularization schemes.

\section{The Nambu-Goto action}

The distance to be considered has the following general form,
\begin{eqnarray}
ds^{2} & = & e^{2A(z)}(dz^{2}+\eta_{ij}dx^{i}dx^{j})\nonumber \\
 & = & G_{\mu\nu}dx^{\mu}dx^{\nu}\,\,\,\,\,\,\,\,\,\,\;\;\mu,\nu=1,\cdots,d\,+1\;\quad.
\end{eqnarray}
It is defined by a metric with no dependence on the boundary coordinates,
which therefore preserves the boundary space Poincaré invariance.
This should be the case if only vacuum properties are considered.
The form of the warp factor $A(z)$ to be considered is,
\begin{equation}
A(z)=-\ln\left(\frac{z}{L}\right)+f(z),\label{eq:warp}
\end{equation}
where $f\left(z\right)$ is a dimensionless function. In this work
$f(z)$ is taken to be a series in even\footnote{Restricting to even powers implies that no odd dimensional condensates
will appear\cite{Quevedo:2013iya}. The motivation for this requirement
is that this is the case in QCD where no odd dimensional condensates
appear.} powers of $z$ , i.e.,
\begin{equation}
f(z)=\sum_{k=1}\alpha_{2k}z^{2k}\;.\label{eq:fz}
\end{equation}
 The case $f\left(z\right)=0$ corresponds to the $AdS$ metric. This
deviation from the AdS case could be produced by a bulk gravity theory
including matter fields. Possible candidates for these bulk gravity
theories have been considered in \cite{Gursoy:2007cb,Goity:2012yj}. 

The area of a surface embedded in this space is given by the NG action,
\begin{equation}
S_{NG}=\frac{1}{2\pi\alpha'}\int d^{2}\sigma\sqrt{g}\,,\label{eq:action}
\end{equation}
where $g$ is the determinant of the induced metric on the surface,
which is given by,
\[
g_{ab}=G_{\mu\nu}\partial_{a}x^{\mu}\partial_{b}x^{\nu}\;,
\]
where $x^{\mu}(a,b)$ are the coordinates of the surface embedded
in the ambient $d+1$ dimensional space. The indices $a,\,b$ refer
to coordinates on the surface.

\section{Rectangular Loop}

The surface contoured by this loop is described by the following embedding,
\begin{eqnarray*}
x^{1} & = & t\hspace{15mm},\;t\in\bigl[{\textstyle {\textstyle -\frac{T}{2},\frac{T}{2}}}\bigr]\\
x^{i} & = & x\hspace{14mm},\;x\in[-a,a]\\
x^{k} & = & 0\hspace{15mm},\,\forall k\neq i\\
x^{5} & = & z=z(x)\,\mathrm{.}
\end{eqnarray*}
the determinat of the induced metric,
\[
g_{ab}=G_{\mu\nu}\partial_{a}x^{\mu}\partial_{b}x^{\nu}\quad(a,b=t,x)
\]
is given by,
\[
\mathrm{det}(g_{ab})=\left[1+z'(x)^{2}\right]\mathrm{e}^{4A(z)}\,,
\]
leading to the following expression for the NG action,
\begin{eqnarray}
S_{NG} & = & \frac{1}{2\pi\alpha'}\int_{-\frac{T}{2}}^{\frac{T}{2}}dt\int_{-a}^{a}dx\,\mathrm{e}^{2A(z(x))}\sqrt{1+z'(x)^{2}}\nonumber \\
 & = & \frac{T}{\pi\alpha'}\int_{0}^{a}dx\,\mathrm{e}^{2A(z(x))}\sqrt{1+z'(x)^{2}}\label{eq: sng_rec}
\end{eqnarray}
where in the last equality traslation and reflection symmetry has
beeen employed. The geometrical setting given above is described in
the following figure,\vspace{2cm}

\begin{figure}[H]
\includegraphics[bb=-1cm 0cm 5cm 0cm,scale=0.7]{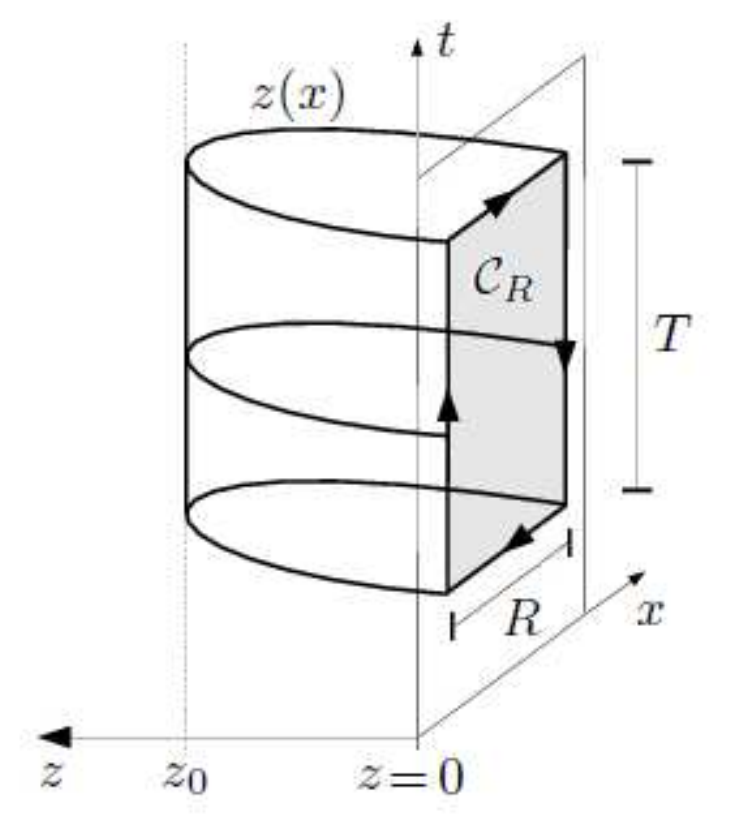}\includegraphics[bb=-6cm 0cm 0cm 5cm,scale=0.7]{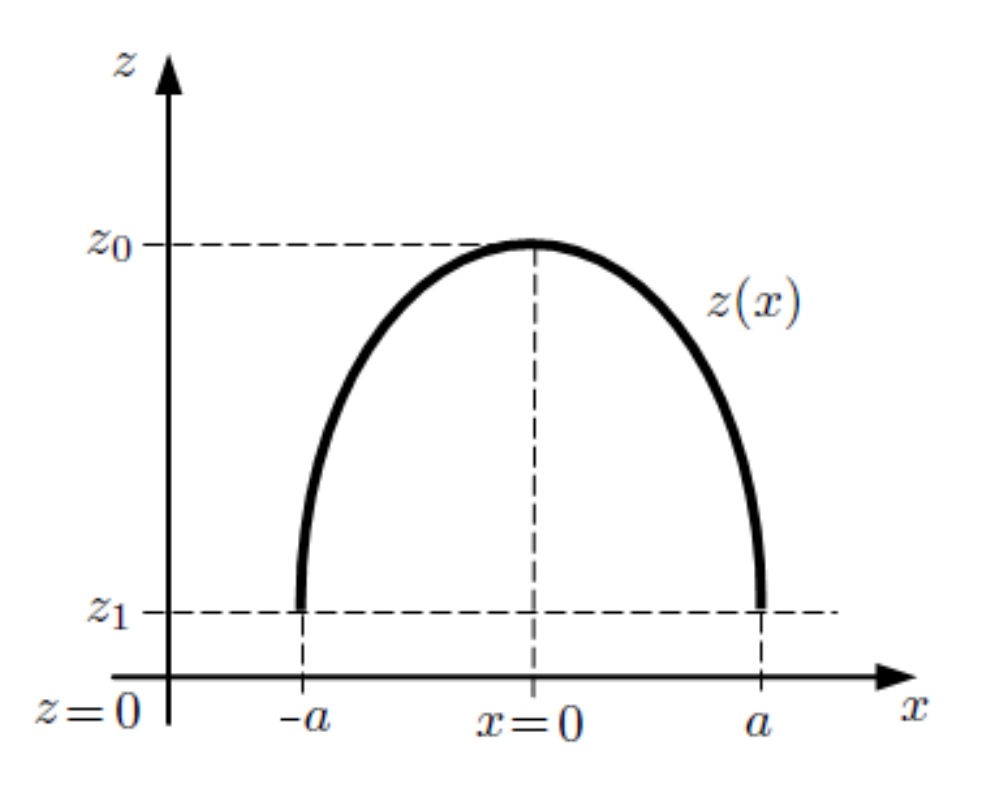}\caption{(a) The rectangular loop is located at $z=0$. The corresponding world
sheet $z(x)$ lives in the bulk. For $T\to\infty$, the world sheet
parametrization $z(x)$ is $t$ independent. (b) A worldsheet section
for fixed $t$. In this case the contour is located at a value $z_{1}$.
This value can be sent to zero after substraction.}
\end{figure}

As described in the second figure, the loop is located at a value
$z_{1}$ of the coordinate orthogonal to the border, in addition the
corresponding minimal surface is required to be regular at the origin.
Therefore the boundary conditions for the minimal surface are,
\[
z(a)=z_{1}\;,z'(0)=0\,\textrm{.}
\]
The potential between static quarks can be obtained from the NG action
as follows,
\begin{equation}
V_{\bar{q}q}(R)=\underset{T\rightarrow+\infty}{\mathrm{lim}}\frac{S_{NG}}{T}=\frac{1}{\pi\alpha'}\int_{0}^{a}dx\,\mathrm{e}^{2A(x)}\sqrt{1+z'(x)^{2}}\,\mathrm{,}\label{eq:iqg}
\end{equation}
where $R=2a$ is the interquark separation.

\subsection{Hamilton-Jacobi approach}

With the boundary conditions mentioned above, the on-shell NG action
is a function of the interquark separarion $a$ and $z_{1}$, the
location of the loop, i.e.,
\[
S_{NG}=S_{NG}(a,z_{1})
\]
the corresponding Hamilton-Jacobi equation is given by,
\[
\frac{\partial S_{NG}(a,z_{1})}{\partial a}+H\left(z_{1},\frac{\partial S_{NG}(a,z_{1})}{\partial z_{1}},a\right)=0\,\mathrm{,}
\]
where $H=H(z,p,x)$ is the Hamiltonian, $p$ the canonical conjugate
momenta to $z$ and $x$ the coordinate along the spatial dimension
of the loop. To make the calculations easier, it is helpfull to neglect
the multiplicative factor $\frac{T}{\pi\alpha'}$ in \ref{eq: sng_rec}
and reintroduce it  in the final expression. Standard methods lead
to,
\begin{eqnarray}
H(z,p,x) & = & p\,z'(z,p,x)-L(z,z'(z,p,x),x)\nonumber \\
 & = & -\frac{\mathrm{e}^{2A(z)}}{\sqrt{1+z^{'2}}}=-\sqrt{\mathrm{e}^{4A(z)}-p^{2}}\,\mathrm{,}\label{eq:H}
\end{eqnarray}
leading to the following form of the HJ equation,
\begin{equation}
\frac{\partial S_{NG}(a,z_{1})}{\partial a}-\sqrt{\mathrm{e}^{4A(z_{1})}-\left[\frac{\partial S_{NG}(a,z_{1})}{\partial z_{1}}\right]^{2}}=0\,\textrm{.}\label{eq:hj-rl}
\end{equation}
In this case, since the lagrangian does not depend on the coordinate
$x$, the Hamiltonian is a constant of motion $E$, thus,
\[
\frac{\partial S_{NG}(a,z_{1})}{\partial a}=\sqrt{\mathrm{e}^{4A(z_{1})}-\left[\frac{\partial S_{NG}(a,z_{1})}{\partial z_{1}}\right]^{2}}=-E
\]
and the value of $E$ can be obtained from (\ref{eq:H}) as follows,
\[
E=-\frac{\mathrm{e}^{2A(z)}}{\sqrt{1+z^{'2}}}=-\mathrm{e}^{2A(z_{0})}
\]
where $z_{0}$ is the maximum value of the coordinate $z$ attained
by the minimal surface, which is therefore such that,
\[
z_{0}=z(0)\:,\,z'(0)=0\,\textrm{.}
\]
An expression for $z_{0}$ as a fucntion of $a$ and $z_{1}$ can
be obtained by means of,
\begin{eqnarray}
a & = & \int_{0}^{a}dx=\int_{z(0)}^{z(a)}\frac{dx}{dz}\,dz\label{eq:z0az1}\\
 & = & \int_{z_{0}}^{z_{1}}\frac{1}{z'}\,dz=\int_{z_{1}}^{z_{0}}\frac{\mathrm{e}^{2A(z_{0})}}{\sqrt{\mathrm{e}^{4A(z)}-\mathrm{e}^{4A(z_{0})}}}\,dz\,\textrm{.}
\end{eqnarray}
 Having a constant of motion, a solution by separation of variables
is possible,
\[
S_{NG}(a,z_{1})=A(a)+Z(z_{1})
\]
replacing in (\ref{eq:hj-rl}) gives,
\begin{eqnarray*}
A'(a) & = & -E\\
Z'(z_{1}) & = & \pm\sqrt{\mathrm{e}^{4A(z_{1})}-E^{2}}\,.
\end{eqnarray*}
The general solution to these equations is\footnote{In the second equation bellow, the minus sign has been chossen. This
choice corresponds to a minimal surface that extends from the border
$z=0$ to greater values of $z$. },
\begin{eqnarray*}
A(a) & = & =-E\cdot a+A_{0}\\
Z(z_{1}) & = & -\int_{z_{inf}}^{z_{1}}\sqrt{\mathrm{e}^{4A(z)}-E^{2}}\,dz\,
\end{eqnarray*}
where the integration constants $A_{0}$ and $z_{inf}$, has to be
determined by choosing adequate boundary conditions. The followign
boundary condition is adopted,
\begin{equation}
\lim_{a\to0}S_{NG}(a,z_{1})=0\;\;,\,\forall\,z_{1}\label{eq:bc}
\end{equation}
this condition is satisfied by the following solution,
\begin{eqnarray*}
A(a) & = & =-E\cdot a\\
Z(z_{1}) & = & -\int_{z_{0}}^{z_{1}}\sqrt{\mathrm{e}^{4A(z)}-E^{2}}\,dz
\end{eqnarray*}
Noting that $\lim_{a\to0}z_{0}=z_{1}$, shows that the required boundary
condition (\ref{eq:bc}) is fullfilled. 

Replacing $S_{NG}(a,z_{1})$ in (\ref{eq:iqg}), the interquark potential
is given by,
\begin{eqnarray}
V_{\bar{q}q}(R) & = & \frac{1}{2\pi\alpha'}\bigg[R\,\mathrm{e}^{2A(z_{0})}+2\int_{z_{1}}^{z_{0}}\sqrt{\mathrm{e}^{4A(z)}-\mathrm{e}^{4A(z_{0})}}\,dz\bigg]\,,\label{eq:iqrl}
\end{eqnarray}
which coincide with the results in \cite{kinar-sonnenschein:1998}.
In order to express this potential in terms of $a$ and $z_{1}$,
equation (\ref{eq:z0az1}) can be employed to obtain $z_{0}$ as a
function of is $a$ and $z_{1}$. In the AdS case $A(z)=-\ln\left(\frac{z}{L}\right)$,
the integrals appearing in (\ref{eq:z0az1}) and (\ref{eq:iqrl})
are elliptic and can be evaluated to give expressions in terms of
the hypergeometric function. In the general case, near the border,
i.e. for $z_{1}\to0$, the integrals can be evaluated up to terms
proportional to positive powers of $z_{1}$, leading to,
\begin{eqnarray}
V_{\bar{q}q}(R) & = & \frac{L^{2}}{2\pi\alpha'}\bigg[\frac{R}{z_{0}^{2}}+\frac{\sqrt{\pi}\,\Gamma(-\frac{1}{4})}{4\,z_{0}\,\Gamma(\frac{5}{4})}+\frac{2}{z_{1}}\bigg]+\mathcal{O}(z_{1}^{3})\nonumber \\
a & = & \frac{\sqrt{\pi}\,z_{0}\,\Gamma(\frac{3}{4})}{\Gamma(\frac{1}{4})}+\mathcal{O}(z_{1}^{3})\,\mathrm{.}\label{sol_rec_ads}
\end{eqnarray}
which clearly shows that there is a divergence for $z_{1}\to0$. This
happens also in the non-AdS case and is related to the divergence
of the metric near the border. A substraction procedure should be
employed to obtain a finite value. This substraction will be discussed
in section 5.

\section{Circular loop}

The surface contoured by the circular loop is described by the following
embedding,
\begin{eqnarray*}
x^{1} & = & 0\\
x^{\sigma} & = & 0\hspace{20mm}\forall k\neq\mu,\nu\\
x^{\mu} & = & r\,\mathrm{cos}(\varphi)\\
x^{\nu} & = & r\,\mathrm{sin}(\varphi)\\
x^{5} & = & z=z(r)\,\;\;,0\leq\varphi\leq2\pi\;,0\leq r\leq a\mathrm{,}
\end{eqnarray*}
it should be noted that the coordiante $z$ has been taken to depend
only on $r$, due to the rotational symmetry of the countour and the
metric. The corresponding geometrical setting is shown in Figure 4.1.
\begin{figure}[H]
\centering{}\includegraphics[scale=0.7]{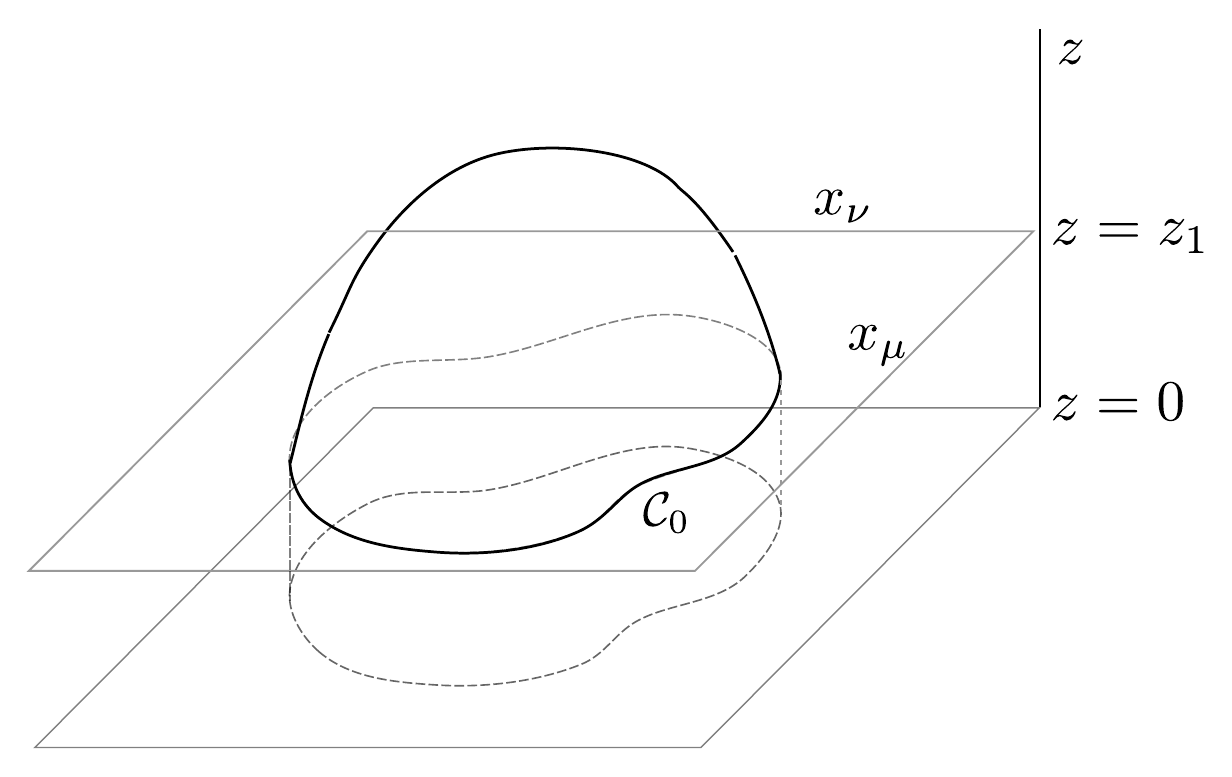}\caption{A loop located at $z=z_{1}$, and the corresponding world sheet.}
\label{z1}
\end{figure}
The induced metric,
\[
g_{ab}=G_{\mu\nu}\partial_{a}x^{\mu}\partial_{b}x^{\nu}\quad(a,b=\varphi,r)
\]
and its determinant are given by,
\[
g_{ab}=\begin{pmatrix}\left[1+z'(r)^{2}\right]\mathrm{e}^{2A(z)} & 0\\
0 & r^{2}\mathrm{e}^{2A(z)}
\end{pmatrix}\;\;\;,\;\;\;\mathrm{det}(g_{ab})=r^{2}\,\left[1+z'(r)^{2}\right]\mathrm{e}^{4A(z)}\,,
\]
which leads to the following expression for the corresponding NG action,
\begin{equation}
S_{NG}=\frac{1}{\alpha'}\int_{0}^{a}dr\,r\,\mathrm{e}^{2A(z)}\sqrt{1+z'(r)^{2}}\,,\label{Sng_r}
\end{equation}
where the $\varphi$ integration has been done, cancelling the $2\pi$
factor in (\ref{eq:action}). It is worth noting that this Lagrangian
depends on the integration variable, and therefore the Hamiltonian
is not a constant of motion in this case. The Euler-Lagrange equations
of motion arising from this action are,
\begin{equation}
r\frac{z''(r)}{1+z'(r)^{2}}+z'(r)-2rA'(z)=0\label{ec-mov}
\end{equation}
the boundary conditions to be considered are,
\begin{equation}
z(a)=z_{1}\;,\,z'(0)=0\label{eq:bcc}
\end{equation}
which correspond to a smooth surface contoured by a circular loop
of radius $a$ located at the value $z_{1}$ of the coordinate $z$
ortogonal to the border. For the AdS case $A(z)=-\mathrm{ln}(\frac{z}{L})$,
the solution to (\ref{ec-mov}) with the boundary conditions (\ref{eq:bcc})
is,
\begin{equation}
z=\sqrt{a^{2}+z_{1}^{2}-r^{2}}\hspace{4mm},0\leq r\leq a\,\mathrm{.}\label{eq:soleqm}
\end{equation}

\subsection{Hamilton-Jacobi approach}

In this case, the NG action is a function of the radius $a$ and the
location $z_{1}$ of the circular loop. The momentum canonically conjugate
to $z$ and the Hamiltonian appearing in the HJ equation are given
by,
\[
p(z,z',r):=\frac{\partial L(z,z',r)}{\partial z'}=\frac{1}{\alpha'}\frac{r\,z'\,\mathrm{e}^{2A(z)}}{\sqrt{1+z^{'2}}}\Rightarrow z'(z,p,a)=\pm\frac{\alpha'\,p}{\sqrt{r^{2}\mathrm{e}^{4A(z)}-\alpha'^{2}\,p^{2}}}\,\mathrm{.}
\]
\begin{eqnarray}
H(z,p,r) & = & p\,z'(z,p,r)-L(z,z'(z,p,r),r)\vspace{3mm}\nonumber \\
 & = & -\frac{p}{z'(z,p,r)}=\mp\frac{1}{\alpha'}\sqrt{r^{2}\,\mathrm{e}^{4A(z)}-\alpha'^{2}p^{2}}\,.\label{H_w}
\end{eqnarray}
Replacing in the HJ equation,
\[
\frac{\partial S_{NG}(a,z_{1})}{\partial a}+H\left(z_{1},\frac{\partial S_{NG}(a,z_{1})}{\partial z_{1}},a\right)=0\,\mathrm{,}
\]
leads to,
\begin{equation}
\frac{\partial S_{NG}(a,z_{1})}{\partial a}\mp\frac{1}{\alpha'}\sqrt{a^{2}\,\mathrm{e}^{4A(z_{1})}-\alpha'^{2}\left[\frac{\partial S_{NG}(a,z_{1})}{\partial z_{1}}\right]^{2}}=0\,\mathrm{,}\label{HJ_w2}
\end{equation}
which implies,
\begin{equation}
\left[\frac{\partial S_{NG}(a,z_{1})}{\partial a}\right]^{2}+\left[\frac{\partial S_{NG}(a,z_{1})}{\partial z_{1}}\right]^{2}=\frac{1}{\alpha'^{2}}a^{2}\,\mathrm{e}^{4A(z_{1})}\,\mathrm{.}\label{HJ}
\end{equation}
In the AdS case this equation is,
\[
\left[\frac{\partial S_{NG}(a,z_{1})}{\partial a}\right]^{2}+\left[\frac{\partial S_{NG}(a,z_{1})}{\partial z_{1}}\right]^{2}=\frac{L^{4}}{\alpha'^{2}}\frac{a^{2}}{z_{1}^{4}}\,\mathrm{,}
\]
whose solution with the boundary condition,
\begin{equation}
\underset{a\rightarrow0^{+}}{\mathrm{lim}}S_{NG}(a,z_{1})\,\equiv0\quad(z_{1}=\mathrm{cte)}\:\mathrm{,}\label{contorno}
\end{equation}
is,
\begin{equation}
S_{NG}^{AdS}(a,z_{1})=\frac{L^{2}}{\alpha'}\left[\sqrt{1+\frac{a^{2}}{z_{1}^{2}}}-1\right]\,\mathrm{.}\label{sol-ads-s}
\end{equation}
which coincides with what is obtained by replacing the solution (\ref{eq:soleqm})
in the NG action (\ref{Sng_r}).

\subsection{Expansion in powers of the radius $a$}

An expansion of the on-shell NG action for the circular loop in powers
of $a$, allows to obtain information about the gluon condensates
in the dual gauge theory \cite{Shifman:1980ui}\cite{Andreev:2007vn}\cite{Quevedo:2013iya}.
It is not totally straightfoward to perform such an expansion. This
can be seem from the result (\ref{sol-ads-s}) for the on-shell NG
action in the AdS case. The series expansion of $S_{NG}^{AdS}(a,z_{1})$
in powers of $a$ is given by,
\begin{equation}
\frac{\alpha'}{L^{2}}\,S_{NG}^{AdS}(a,z_{1})=\frac{a^{2}}{2z_{1}^{2}}-\frac{a^{4}}{8z_{1}^{4}}+\frac{a^{6}}{16z_{1}^{6}}+\mathcal{O}(a^{7})\label{eq:expa}
\end{equation}
which is convergent for $\frac{a}{z_{1}}<1$. Therefore such an expansion
is not suited to reproduce the behaviour of $S_{NG}^{AdS}(a,z_{1})$
for $z_{1}\to0$ and $a$ fixed. Indeed, (\ref{sol-ads-s}) shows
that, 
\begin{equation}
\frac{\alpha'}{L^{2}}\,S_{NG}^{AdS}(a,z_{1})\overset{_{z_{1}\ll1}}{=}\frac{a}{z_{1}}\label{eq:divads}
\end{equation}
In this respect it is convenient to consider the NG action in terms
of the variables $w_{1}=\frac{z_{1}}{a}$ and $a$ instead of $z_{1}$
and $a$. Doing this for the AdS case gives,
\[
\frac{\alpha'}{L^{2}}\,S_{NG}^{AdS}(a,w_{1}a)=\frac{\sqrt{1+w_{1}^{2}}}{w_{1}}-1\:\mathrm{,}
\]
whose Laurent expansion for $w_{1}\ll1$ reproduces the divergence
term $1/w_{1}$ in (\ref{eq:divads}), this is not the case for the
expansion (\ref{eq:expa}).

Defining the action $S(a,w_{1})$ by,
\[
S(a,w_{1})=S_{NG}(a,w_{1}a)
\]
and taking into account that,

\begin{eqnarray*}
\frac{\partial S_{NG}(a,z_{1})}{\partial a} & = & \frac{\partial S(a,w_{1})}{\partial a}-\frac{w_{1}}{a}\,\frac{\partial S(a,w_{1})}{\partial w_{1}}\\[0.4em]
\frac{\partial S_{NG}(a,z_{1})}{\partial z_{1}} & = & \frac{1}{a}\,\frac{\partial S(a,w_{1})}{\partial w_{1}}\,\mathrm{,}
\end{eqnarray*}
the HJ equation is rewritten as follows,
\begin{eqnarray}
a^{2}\,\left[\frac{\partial S(a,w_{1})}{\partial a}\right]^{2}+\left(1+w_{1}^{2}\right)\left[\frac{\partial S(a,w_{1})}{\partial w_{1}}\right]^{2}\nonumber \\[0.6em]
-2\,w_{1}\,a\,\frac{\partial S(a,w_{1})}{\partial a}\,\frac{\partial S(a,w_{1})}{\partial w_{1}} & = & \frac{1}{\alpha'^{2}}\,a^{4}\,\mathrm{e}^{4A(w_{1}\cdot a)}\,\mathrm{,}\label{HJ2}
\end{eqnarray}
the boundary condition (\ref{contorno}) is now,
\begin{eqnarray}
0=\underset{a\rightarrow0^{+}}{\mathrm{lim}}S_{NG}\left(a,z_{1}\right)=\begin{split}\underset{a\rightarrow0^{+}}{\mathrm{lim}}S\left(a,\frac{z_{1}}{a}\right)\end{split}
 & \quad(z_{1}=\mathrm{cst.)\:\mathrm{.}}\label{contorno2}
\end{eqnarray}
Next the following power series expansion is considered,
\begin{eqnarray}
S\left(a,w_{1}\right) & = & \frac{L^{2}}{\alpha'}\,\sum_{n=0}^{\infty}s_{2n}\left(w_{1}\right)\,a{}^{2n}\:\mathrm{,}\label{ansatz}
\end{eqnarray}
replacing this expansion in (\ref{HJ2}) leads to,
\begin{eqnarray}
\sum_{n=0}^{\infty}\left(\sum_{k=0}^{n}4(k+1)(n-k+1)\;s_{2(k+1)}(w_{1})\;s_{2(n-k+1)}(w_{1})\right)\,a^{2n+4}-\nonumber \\
2\,w_{1}\,\sum_{n=0}^{\infty}\left(\sum_{k=0}^{n}2(k+1)\;s_{2(k+1)}(w_{1})\;s'_{2(n-k)}(w_{1})\right)a^{2n+2}+\nonumber \\
\left(1+w_{1}^{2}\right)\sum_{n=0}^{\infty}\left(\sum_{k=0}^{n}s'_{2k}(w_{1})\;s'_{2(n-k)}(w_{1})\right)\,a^{2n}-\sum_{n=0}^{\infty}\beta_{2n}\,w_{1}^{2n-4}\,a^{2n} & = & 0\,,\label{eq:eqexpan}
\end{eqnarray}
where $\beta_{m}$ are the power series expansion coefficients of
$\mathrm{e}^{4A(w_{1}a)}$, i.e.,
\[
\mathrm{e}^{4A(w_{1}a)}=\frac{L^{4}}{\left(w_{1}a\right)^{4}}\,\mathrm{e}^{4f(w_{1}a)}=\frac{L^{4}}{\left(w_{1}a\right)^{4}}\,\sum_{n=0}^{\infty}\beta_{2n}\left(w_{1}a\right)^{2n}\,,
\]
these coefficients can be written as polynomials in the $\alpha$
coefficients appearing in (\ref{eq:fz}). Equating to zero the coefficient
of $a^{n}$ in the l.h.s. of (\ref{eq:eqexpan}), leads to,
\begin{eqnarray}
2\left(1+w_{1}^{2}\right)\,s'_{0}(w_{1})\,s'_{2n}(w_{1})\,-4n\,w_{1\,}s'_{0}(w_{1})\,s_{2n}(w_{1})\,+\nonumber \\
\sum_{k=1}^{n-1}\left\{ \left(1+w_{1}^{2}\right)\,s'_{2k}(w_{1})\,s'{}_{2(n-k)}(w_{1})\,+4k\left(n-k\right)\,s{}_{2k}(w_{1})\,s{}_{2(n-k)}(w_{1})\right.\label{HJ4}\\
\left.-4w_{1}\,k\,s{}_{2k}(w_{1})\,s'{}_{2(n-k)}(w_{1})\right\} -\beta_{2n}\,w_{1}^{2n-4} & = & 0\,,\nonumber 
\end{eqnarray}
valid for $n=0,1,2,\cdots$. The boundary condition (\ref{contorno2})
leads to, 
\begin{equation}
\underset{a\rightarrow0^{+}}{\mathrm{lim}}S\left(a,z_{1}/a\right)=0\,\Longleftrightarrow\underset{a\rightarrow0^{+}}{\mathrm{lim}}s_{2n}\left(z_{1}/a\right)\cdot a^{2n}=0\;\;\forall\,n,\,z_{1}=\textrm{cst}.\label{eq:bc1}
\end{equation}
For a given $n$, equation (\ref{HJ4}) involves the functions $s_{2k}(w_{1})$
and $s'_{2k}(w_{1})$ for $0\leq k\leq n$. Therefore starting with
$n=0$, the resulting equation only involves $s_{0}(w_{1})$ and $s'_{0}(w_{1})$,
solving for them they can be replaced in the equation for $n=1$,
to get $s_{2}(w_{1})$ and $s'_{2}(w_{1})$ and so on. The equation
for $n=0$ and its solution satisfying the boundary condition (\ref{eq:bc1})
are,
\begin{equation}
[s'_{0}(w_{1})]^{2}=\frac{1}{w_{1}^{4}\left(1+w_{1}^{2}\right)}\:\implies\:s{}_{0}(w_{1})=+\left(\frac{\sqrt{1+w_{1}^{2}}}{w_{1}}-1\right)\:\mathrm{,}\label{s0}
\end{equation}
where the sign in the second equation has been choosen so as to get
a positive area for non-vanishing radius. For the cases with $n=1,2,\cdots$
the differential equations to be considered are of the form,
\begin{equation}
A^{(2n)}(w_{1})\,s'_{2n}(w_{1})\,+B^{(2n)}(w_{1})\,s_{2n}(w_{1})\,+\,C^{(2n)}(w_{1})=0\,\textrm{,}\label{HJ5}
\end{equation}
the general solution to this equation is,
\begin{eqnarray}
s_{2n}(w_{1}) & = & c^{(2n)}\,\mathrm{e}^{F(w_{1})}-\mathrm{e}^{F(w_{1})}\,\int_{0}^{w_{1}}\mathrm{e}^{-F(x)}\,\frac{C^{(2n)}(x)}{A^{(2n)}(x)}\,\mathrm{d}x\,\textrm{,}\nonumber \\
F(w_{1}) & = & -\int_{0}^{w_{1}}\frac{B^{(2n)}(x)}{A^{(2n)}(x)}\,\mathrm{d}x\,\textrm{,}\label{sol}
\end{eqnarray}
where $c^{(2n)}$ is a constant to be determined using the boundary
condition (\ref{eq:bc1}). Eq. (\ref{HJ4}) implies that,
\begin{eqnarray*}
\frac{B^{(2n)}(x)}{A^{(2n)}(x)} & =-2n & \frac{w_{1}}{1+w_{1}^{2}}\quad\Longrightarrow\quad F(w_{1})=\mathrm{ln}\left[\left(1+w_{1}^{2}\right)^{n}\right]\,\textrm{,}
\end{eqnarray*}
replacing in (\ref{sol}) leads to,
\begin{eqnarray}
s_{2n}(w_{1}) & = & c^{(2n)}\,\left(1+w_{1}^{2}\right)^{n}+\frac{1}{2}\,\left(1+w_{1}^{2}\right)^{n}\,\int_{0}^{w_{1}}\frac{x^{2}}{\left(1+x^{2}\right)^{n+\frac{1}{2}}}\,C^{(2n)}(x)\,\mathrm{d}x\,\textrm{,}\label{sol2}
\end{eqnarray}
where the following equalities were employed $A^{(2n)}(x)=2\left(1+x^{2}\right)\,s'_{0}(x)=-\frac{2\sqrt{1+x^{2}}}{x^{2}}$.
Inposing the boundary condition (\ref{eq:bc1}) leads to,
\[
c^{(2n)}=-\frac{1}{2}\int_{0}^{+\infty}\frac{x^{2}}{\left(1+x^{2}\right)^{n+\frac{1}{2}}}\,C^{(2n)}(x)\,\mathrm{d}x\,\textrm{,}
\]
which replacing in (\ref{sol2}) gives,
\begin{equation}
s_{2n}(w_{1})=-\frac{1}{2}\,\big(1+w_{1}^{2}\big)\,\int_{w_{1}}^{+\infty}\frac{x^{2}}{\big(1+x^{2}\big)^{n+\frac{1}{2}}}\,C^{(2n)}(x)\,\mathrm{d}x\,\mathrm{,}\label{sol3}
\end{equation}
the functions $C^{(2n)}$ appearing in this expression are obtained
form (\ref{HJ4}),
\begin{eqnarray}
C^{(2n)}(x) & = & \sum_{k=1}^{n-1}\left\{ \left(1+x{}^{2}\right)\,s'_{2k}(x)\,s'{}_{2(n-k)}(x)\,+\,4k\left(n-k\right)\,s{}_{2k}(x)\,s{}_{2(n-k)}(x)\right.\nonumber \\[0.3em]
 &  & \left.-4w_{1}\,k\,s{}_{2k}(x)\,s'{}_{2(n-k)}(x)\right\} -\beta_{2n}\,x^{2n-4}\,\textrm{.}\label{C2n}
\end{eqnarray}
for $n=0,1,2,3$ the results for $C^{(2n)}(x)$ and $s_{2n}(w_{1})$
are given in appendix A.

\section{Substraction\label{sec:Substraction}}

\begin{figure}[H]
\begin{centering}
\includegraphics[width=8cm]{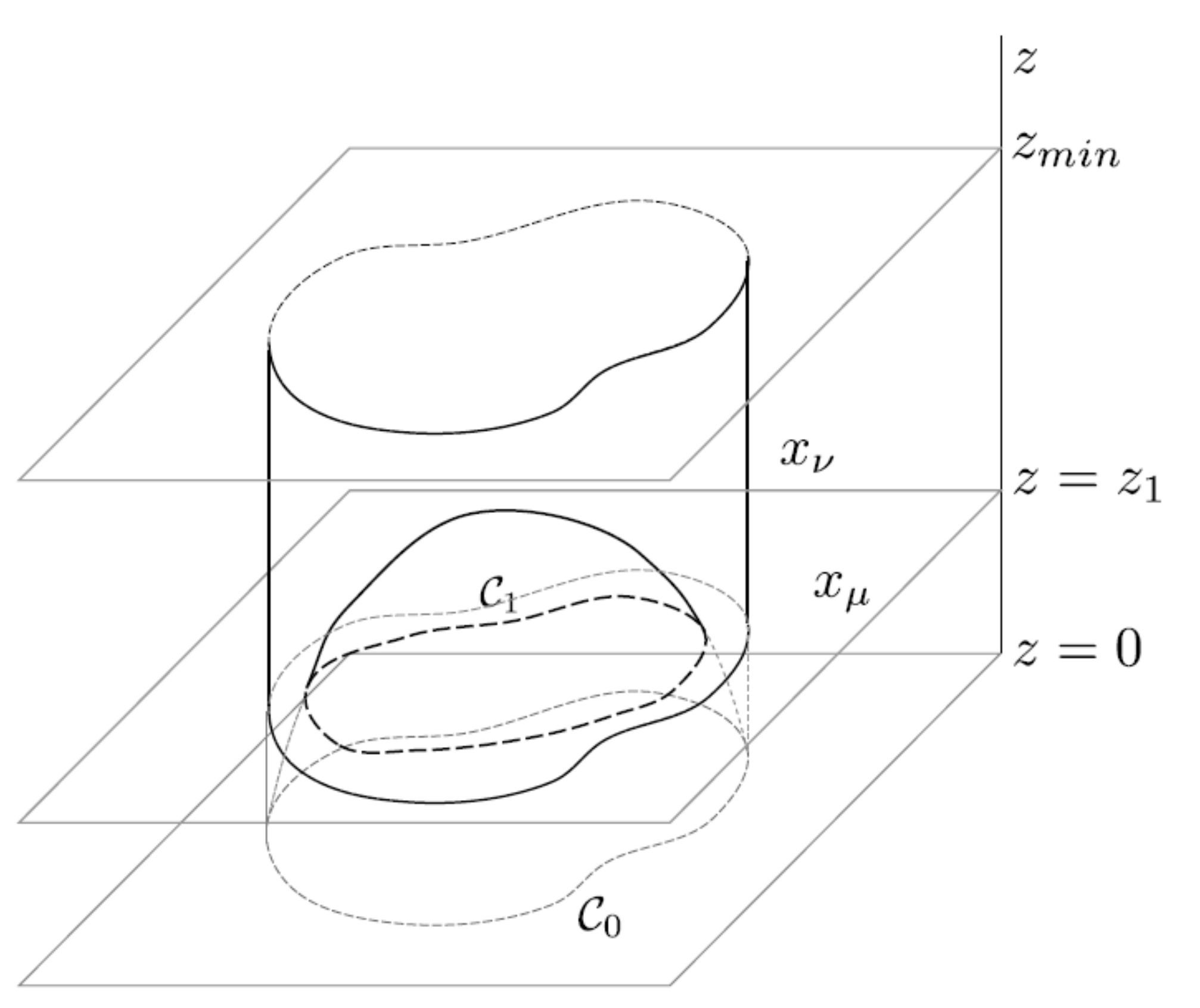}
\par\end{centering}

\caption{Substraction scheme.\label{fig:Substraction-schema.}}
\label{subs}
\end{figure}

The substraction procedure employed is essentially the same as the
one in \cite{Maldacena:1998im}. It has been extended and applied
to the non-AdS case in \cite{Quevedo:2013iya}. The subtraction $S_{CT}$
to the NG action has a clear geometrical meaning which is illustrated
in fig. \ref{subs}.

It corresponds to the area of a cylinder with section given by a countour
such that the minimal area surrounded by this countour at $z=0$ intersects
the plane $z=z_{1}$ with the original countour. For the case of confining
warp factors the extension of this cylinder in the $z$ direction
is regulated by an infrared cut-off $z_{IR}$. This is so because
for those geometries the warp factor necesarily presents a minimum
$z_{m}$ above which the warp factor grows\cite{kinar-sonnenschein:1998}.
In \cite{Quevedo:2013iya} it is argued that a natural candidate for
this infrared scale is the location of the warp factor minimum $z_{m}$.
In any case as we shall see bellow the physical quantities to be calculated
do not depend on this scale.

For the case of the rectangular loop $S_{CT}$ is given by,
\[
S_{CT}=\frac{2}{2\pi\alpha'}\int_{-\frac{T}{2}}^{\frac{T}{2}}dt\int_{z_{1}}^{z_{IR}}dz\,\mathrm{e}^{2A(z)}
\]
for the AdS case this subtraction is,
\[
S_{CT}^{AdS}=\frac{T}{\pi\alpha'}\int_{z_{1}}^{z_{IR}}dz\,\frac{1}{z^{2}}=\frac{T}{\pi\alpha'}\frac{1}{z_{1}}-\frac{T}{\pi\alpha'}\frac{1}{z_{IR}}
\]
which when substracted to $S_{NG}$ cancells the divergent term in
(\ref{sol_rec_ads}). 

For the circular loop the substracted on-shell NG action is given
by,
\begin{equation}
S_{NG}^{sub}=S_{NG}-S_{CT}=S_{NG}-\frac{1}{2\pi\alpha'}\,r_{0}(a,z_{1})\int_{z_{1}}^{z_{IR}}dz\,e^{2A(z)}\,,\label{eq:nambu-goto substraida}
\end{equation}
 For the AdS case the function $r_{0}(a,z_{1})$ is fixed by conformal
invariance and is given by,
\begin{equation}
r_{0}^{AdS}(a,z_{1})=2\pi\sqrt{a^{2}+z_{1}^{2}}\,,\label{eq:rads}
\end{equation}
leading to,
\begin{equation}
S_{NG}^{sub,AdS}=-\frac{L^{2}}{\alpha'}\,.\label{eq:NGadssubs}
\end{equation}

For the non-AdS case one could take $r_{0}(a,z_{1})$ to be the radius
of a loop located at the boundary whose minimal surface would intersect
the plane $z=z_{1}$ with a circle of radius $a$. However as explained
in \cite{Quevedo:2013iya} it is simpler to take the AdS expression
given by (\ref{eq:rads}), which presents no conflict with conformal
invariance and leads to a finite substrated NG action even in the
non-AdS case.

The choice of $z_{IR}$ does not affect the result for the condensates,
since it only affects the coefficient of the perimeter in the expansion
of the on-shell NG action in powers of the radius $a$.

\subsection{Computing the subtraction}

Fot the circular loop, in terms of $w_{1}=z_{1}/a$, the subtraction
is given by,
\[
S_{CT}=\frac{1}{\alpha'}\,a\sqrt{1+w_{1}^{2}}\int_{w_{1}\cdot a}^{z_{IR}}\mathrm{e}^{2A\left(z\right)}\,\mathrm{d}z\,
\]
the warp factor to be considered is given by (\ref{eq:fz}), i.e.,
\[
A(z)=-\log\left(\frac{z}{L}\right)+f(z)\;\;,\,f(z)=\sum_{k=1}\alpha_{2k}z^{2k}
\]
thus,
\begin{eqnarray*}
\frac{\alpha'}{L^{2}}\,S_{CT} & = & \frac{1}{L^{2}}\,a\sqrt{1+w_{1}^{2}}\int_{w_{1}\cdot a}^{z_{IR}}\mathrm{e}^{-2\log(\frac{z}{L})+2f(z)}\,\mathrm{d}z=\frac{1}{L^{2}}\,a\sqrt{1+w_{1}^{2}}\int_{w_{1}\cdot a}^{z_{IR}}\frac{L^{2}}{z^{2}}\mathrm{e}^{2f(z)}\,\mathrm{d}z\\
 & = & a\sqrt{1+w_{1}^{2}}\int_{w_{1}\cdot a}^{z_{IR}}\frac{1}{z^{2}}\left(\mathrm{1+e}^{2f(z)}-1\right)\,\mathrm{d}z\\
 & = & a\sqrt{1+w_{1}^{2}}\int_{w_{1}\cdot a}^{z_{IR}}\frac{1}{z^{2}}\,\mathrm{d}z\,+a\sqrt{1+w_{1}^{2}}\int_{w_{1}\cdot a}^{z_{IR}}\frac{\mathrm{e}^{2f(z)}-1}{z^{2}}\,\mathrm{d}z\\
 & = & a\sqrt{1+w_{1}^{2}}\left.\left(-\frac{1}{z}\right)\right|_{w_{1}\cdot a}^{z_{IR}}\,+a\sqrt{1+w_{1}^{2}}\int_{w_{1}\cdot a}^{z_{IR}}\frac{\mathrm{e}^{2f(z)}-1}{z^{2}}\,\mathrm{d}z\\
 & = & \frac{\sqrt{1+w_{1}^{2}}}{w_{1}}+a\sqrt{1+w_{1}^{2}}\left[-\frac{1}{z_{IR}}+\int_{w_{1}\cdot a}^{z_{IR}}\frac{\mathrm{e}^{2f(z)}-1}{z^{2}}\,\mathrm{d}z\right]\,,
\end{eqnarray*}
the integrand in this last equation has no singularities in the integration
region. Therefore the only singular term of this expression for $w_{1}\to0$
is the first. This singular part coincides with the one in the AdS
case. Indeed it is produced by the AdS term $-\log\left(\frac{z}{L}\right)$
of the warp factor. Thus the singular part of the counterterm is not
affected by the addition of $f(z)$ to the warp factor.

\subsection{The subtracted NG action}

For the rectangular loop, the substracted NG action is given by,

\begin{eqnarray}
S_{NG}^{sub} & = & S_{NG}-S_{CT}\nonumber \\
 & = & \frac{T}{2\pi\alpha'}\bigg[R\,\mathrm{e}^{2A(z_{0})}+2\int_{z_{1}}^{z_{0}}\sqrt{\mathrm{e}^{4A(z)}-\mathrm{e}^{4A(z_{0})}}\,dz\bigg]-\frac{T}{\pi\alpha'}\int_{z_{1}}^{z_{IR}}\mathrm{e}^{2A(z)}\,dz\nonumber \\
 & = & \frac{T}{2\pi\alpha'}\bigg[R\,\mathrm{e}^{2A(z_{0})}+2\int_{z_{1}}^{z_{0}}\sqrt{\mathrm{e}^{4A(z)}-\mathrm{e}^{4A(z_{0})}}-\mathrm{e}^{2A(z)}\,dz\bigg]\nonumber \\
 &  & -\frac{T}{\pi\alpha'}\int_{z_{0}}^{z_{IR}}\mathrm{e}^{2A(z)}\,dz\label{eq:ssub_rect}
\end{eqnarray}
The first integral is now finite even when $z_{1}\to0$. This can
be seen by noting that the integrand has no singularities and is well
behaved when $z\to0$, this is shown bellow,
\begin{eqnarray*}
\sqrt{\mathrm{e}^{4A(z)}-\mathrm{e}^{4A(z_{0})}}-\mathrm{e}^{2A(z)} & = & \sqrt{\mathrm{e}^{-4\log(\frac{z}{L})+4f(z)}-\mathrm{e}^{-4\log(\frac{z_{0}}{L})+4f(z_{0})}}-\mathrm{e}^{-2\log(\frac{z}{L})+2f(z)}\\
 & = & \sqrt{\frac{L^{4}}{z^{4}}\mathrm{e}^{4f(z)}-\frac{L^{4}}{z_{0}^{4}}\mathrm{e}^{4f(z_{0})}}-\frac{L^{2}}{z^{2}}\mathrm{e}^{2f(z)}\\
 & = & L^{2}\mathrm{e}^{2f(z)}\,\frac{\sqrt{1-\left(\frac{z}{z_{0}}\right)^{4}\mathrm{e}^{4\left[f(z_{0})-f(z)\right]}}-1}{z^{2}}\\
 & = & L^{2}\mathrm{e}^{2f(z)}\,\frac{-\frac{1}{2}\left(\frac{z}{z_{0}}\right)^{4}\mathrm{e}^{4\left[f(z_{0})-f(z)\right]}+\mathcal{O}\left(\left(\frac{z}{z_{0}}\right)^{4}\mathrm{e}^{4\left[f(z_{0})-f(z)\right]}\right)^{2}}{z^{2}}
\end{eqnarray*}
where in the last step the power series expansion of the square root
was employed. The second integral in (\ref{eq:ssub_rect}) is convergent
and depends on the infrared cutoff $z_{IR}$. It is shown in \cite{kinar-sonnenschein:1998}
that when the interquark separation is big ($R=2a\gg L$) the value
of $z_{0}$ goes to $z_{m}$, the minimum of the warp factor, and
consequently the interquark potential has a linear dependence on $R$,

\begin{eqnarray*}
V_{\bar{q}q}(R) & = & \sigma\,R+V_{0}\quad\textrm{if}\:R\gg L
\end{eqnarray*}
where $\sigma=\frac{\mathrm{e}^{2A(z_{m})}}{2\pi\alpha^{'}}$ is the
quark-antiquark string tension. Warp factors such that $\mathrm{e}^{2A(z_{m})}\neq0$
predict linear confinement as happens in QCD.\\

For the circular loop, according to (\ref{eq:nambu-goto substraida}),
the subtracted NG action $S_{NG}^{sub}$ can be written as follows,
\begin{eqnarray*}
\frac{\alpha'}{L^{2}}\,S_{NG}^{sub}(a,w_{1}) & = & \frac{\alpha'}{L^{2}}\,S_{NG}(a,w_{1})-\frac{\alpha'}{L^{2}}\,S_{CT}(a,w_{1})\\[0.4em]
 & = & -1+\sum_{n=1}^{\infty}s_{2n}\left(w_{1}\right)\,a{}^{2n}-a\sqrt{1+w_{1}^{2}}\left[-\frac{1}{z_{IR}}+\int_{w_{1}\cdot a}^{z_{IR}}\frac{\mathrm{e}^{2f(z)}-1}{z^{2}}\,\mathrm{d}z\right]\\
 & \underset{w_{1}\rightarrow0}{\longrightarrow} & -1+\varPhi\left(z_{IR},\left\{ \alpha_{2n}\right\} _{n=1}^{\infty}\right)\,a+\sum_{n=1}^{\infty}s_{2n}(0)\,a{}^{2n}\,\textrm{,}
\end{eqnarray*}
where,
\begin{eqnarray*}
\varPhi\left(z_{IR},\left\{ \alpha_{2n}\right\} _{n=1}^{\infty}\right) & := & \frac{1}{z_{IR}}-\int_{0}^{z_{IR}}\frac{\mathrm{e}^{2f(z)}-1}{z^{2}}\,\mathrm{d}z\,\textrm{.}
\end{eqnarray*}
 In the, limit $w_{1}\to0$ with $a$ fixed the result for substracted
on-shell NG action up to order $a^{6}$ is,
\begin{eqnarray}
\frac{\alpha'}{L^{2}}\,S_{NG}^{sub}\left(a\,,\,z_{1}=0\right) & = & -1+\varPhi\,a+2\alpha_{2}\,a^{2}+\left[\left(\frac{34}{3}-8\,\log(4)\right)\alpha_{2}^{2}+\frac{2}{3}\,\alpha_{4}\right]\,a^{4}\:\nonumber \\
 &  & +\frac{2}{45}\big[\left(1774-2280\,\log\left(4\right)+720\,\log^{2}\left(4\right)\right)\alpha_{2}^{3}\vspace{3mm}\label{solusub}\\[0.4em]
 &  & +\left(326-240\,\log\left(4\right)\right)\alpha_{2}\,\alpha_{4}+9\,\alpha_{6}\big]\,a^{6}+\ldots\mathrm{.}\nonumber 
\end{eqnarray}
Taking $\alpha_{2}=0$ (which corresponds to the absence of a dimension
$2$ condensate in the border gauge theory) the calculation can be
extended to higher orders. The result in this case up to order $a^{10}$
is,
\begin{eqnarray}
\frac{\alpha'}{L^{2}}\,S_{NG}^{sub}\left(a\,,\,z_{1}=0\right) & = & -1+\left.\varPhi\right|_{\alpha_{2}=0}\,a+\frac{2}{3}\,\alpha_{4}\,a^{4}\,+\frac{2}{5}\,\alpha_{6}\,a^{6}\nonumber \\[0.4em]
 &  & +\left[\left(\frac{4222}{945}-\frac{32\,\log(4)}{9}\right)\alpha_{4}^{2}+\frac{2}{7}\alpha_{8}\right]\,a^{8}\label{solusub2}\\[0.5em]
 &  & +\left[\frac{4\left(2999-2520\,\log(4)\right)}{1575}\alpha_{4}\alpha_{6}+\frac{2}{9}\alpha_{10}\right]\,a^{10}+\ldots\mathrm{.}\nonumber 
\end{eqnarray}

\section{Comparison with other computations}

In this section the computations done above are compared with the
analogous ones computed using the $\epsilon$-scheme. This is particularly
relevant since, although the results coincide for the case of the
rectangular loop, this is not the case for the circular loop. The
result for the expansion coefficients $s_{2n}$ of the substracted
NG action in powers of the radius $a$ in the limit $z_{1}\to0$ for
the case of the circular loop do not coincide between both schemes.
The source of this coincidence and discrepancy are analysed below.
In order to do this it is necesary to consider the process of regularization/substraction
involved in the $\epsilon$-scheme. The $\epsilon$-regularization
scheme is despicted in the following figure,
\begin{figure}[H]
\centering{}\includegraphics[scale=0.6]{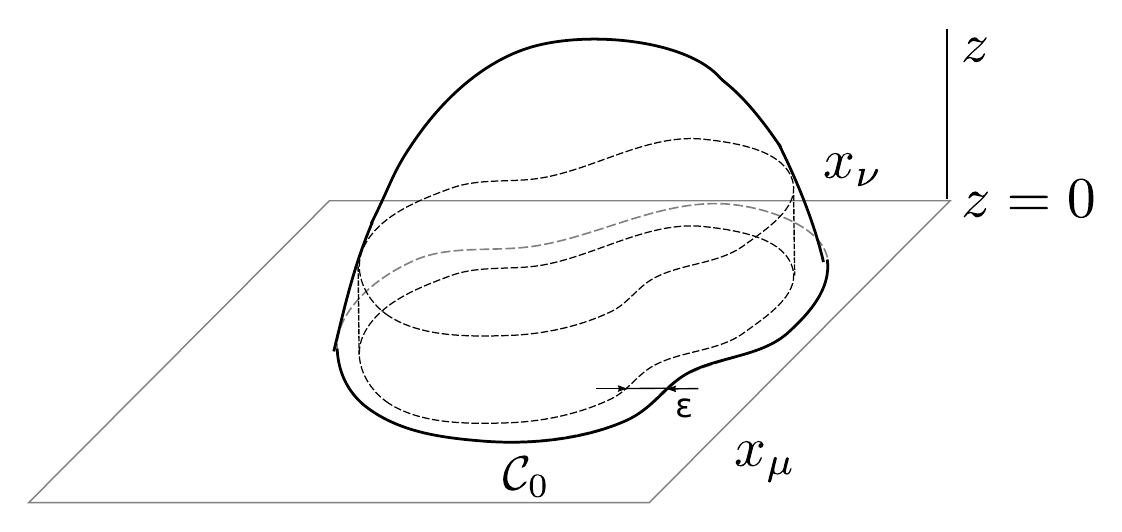}\caption{The $\epsilon$ regularization. .}
\end{figure}
For the non-conformal case this scheme is considered in \cite{Andreev:2007vn}.
It consists in locating the countour of the loop at the boundary $z=0$,
obtaining the corresponding minimal surface and computing its area
up to a value of the coordinate orthogonal to the border which is
an amount $\epsilon$ less than the one of the original countour.
Thus varying the diameter of the countour or the cut-off $\epsilon$
amounts to the same thing for the on-shell NG action. In addition,
the boundary conditions are given at a value of $z$ that does not
correspond to the location of the base of the surface whose area is
calculated. In other words, the value of the $z$ coordinate corresponding
to the location of the loop countour is not a variable and is fixed
to zero. Therefore this scheme is not well suited for employing the
HJ method. Nevertheless, the value of the on-shell NG action can be
calculated by solving the equation of motion (\ref{ec-mov}) with
boundary conditions,
\[
z(a)=0\;,\,z'(0)=0
\]
and replacing in the NG action. Alternativedly, one can employ the
solution to the equation of motion (\ref{ec-mov}) with the boundary
conditions (\ref{eq:bcc}) and take the limit $z_{1}\to0$ in the
integrand of the NG action. This last approach was employed in \cite{Quevedo:2013iya}.

The substraction to be employed is the same as the one described in
section \ref{sec:Substraction}.

\subsection{The rectangular loop}

The solution to the equations of motion for this case can be obtained
by noting that the Hamiltonian corresponding to the Lagrangian appearing
in (\ref{eq: sng_rec}) is a constant of motion $E$, given by,
\[
E=-\frac{e^{2A}}{\sqrt{1+z'(x)^{2}}}
\]
where the $'$ indicates derivative respect to $x$. This leads to
the following linear ordinary diferential equation,
\[
z'(x)=\sqrt{\frac{e^{4A}}{E^{2}}-1}
\]
from which it is simple to obtain $x$ as a function of $z$,

\begin{equation}
x(z)=a-\int_{z_{1}}^{z}dz\frac{E}{\sqrt{e^{4A}-E^{2}}}\label{eq:solxz}
\end{equation}
this solution satisfies the boundary condition $x(z_{1})=a$, and
therefore corresponds to a pair of static quarks located at $z=z_{1}$
and separated a distance $2a$. The relation between $z_{1}$, $z_{0}$,
$E$ and $a$ can be obtained using that by definition $x(z_{0})=0$,
leading to,
\begin{equation}
a=\int_{z_{1}}^{z_{0}}dz\frac{E}{\sqrt{e^{4A}-E^{2}}}\label{eq:az0z1}
\end{equation}

In order to insert this solution in the Nambu-Goto action it is convenient
to consider an alternative embedding of the same surface in the five
dimensional space. In this embedding one considers $x$ as a function
of $z$. It is given by,
\begin{eqnarray*}
x^{1} & = & t\hspace{15mm},\;t\in\bigl[{\textstyle {\textstyle -\frac{T}{2},\frac{T}{2}}}\bigr]\\
x^{i} & =x= & x(z)\qquad\;,\;x\in[-a,a]\\
x^{k} & = & 0\hspace{15mm},\,\forall k\neq i\\
x^{5} & = & z\,\mathrm{.}
\end{eqnarray*}
the NG action is given in terms of this embedding by,
\begin{equation}
S_{NG}=\frac{1}{2\pi\alpha'}\int_{-\frac{T}{2}}^{\frac{T}{2}}dt\int_{z_{1}}^{z_{0}}dz\,\mathrm{e}^{2A(z)}\sqrt{1+x'(z)^{2}}\label{eq:ngxz}
\end{equation}
where now the $'$ indicates derivative respect to $z$, $z_{1}$
is the location of the loop countour and $z_{0}$ the maximum value
of $z$ attained by the minimal surface. Evaluation of the NG action
in the solution of the equations of motion amounts to replace (\ref{eq:solxz})
in (\ref{eq:ngxz}). The integrand in (\ref{eq:ngxz}) is independent
of $z_{1}$, because $x'(z)$ is independent of $z_{1}$ as implied
by (\ref{eq:solxz}). The lower integration limit is common to the
HJ and the $\epsilon$-regularization. The upper limit of integration
is to be determined as a function of $a$ and $z_{1}$ by means of
(\ref{eq:az0z1}). For the $\epsilon$-regularization this amounts
to take $z_{1}\to0$ in (\ref{eq:az0z1}). The function $z_{0}(a,z_{1})$
just gives the maximum of the minimal surface, this function has no
singularities for any value of $z_{1}\geq0$. Therefore taking the
limit $z_{1}\to0$, before or after the integration makes no difference.
Thus the result for the substracted NG action is the same for both
schemes.

\subsection{Circular loop}

The results for the expansion coefficients $s_{2n}$ of the NG action
in powers of the radius $a$ for the case of the circular loop are
shown in the table bellow. The first column corresponds to the results
computed with HJ-scheme. The second column corresponds to the results
computed in \cite{Quevedo:2013iya} using the $\epsilon$-scheme .
The first two rows in column $\epsilon$ also coincide with the results
in \cite{Andreev:2007vn}, which takes $\alpha_{n}=\delta_{n2}\alpha_{2}$.
In the $\epsilon$-computation the loop is located at $z=0$. 
\begin{table}[H]
\begin{tabular}{|c|c|c|}
\hline 
$s_{2n}$ & HJ & $\epsilon$\tabularnewline
\hline 
\hline 
$s_{2}$ & $2\alpha_{2}$ & $\frac{10}{3}\alpha_{2}$\tabularnewline
\hline 
$s_{4}$ & $\frac{2}{3}(17-24\log2)\alpha_{2}^{2}+\frac{2}{3}\alpha_{4}$ & $\frac{14}{9}(17-24\log2)\alpha_{2}^{2}+\frac{14}{9}\alpha_{4}$\tabularnewline
\hline 
$s_{6}$ & $\frac{2}{5}\alpha_{6}$ & $\frac{3}{5}\alpha_{6}$\tabularnewline
\hline 
$s_{8}$ & $\frac{2}{945}(2111-3360\log2)\alpha_{4}^{2}+\frac{2}{7}\alpha_{8}$ & $\frac{11}{5670}[(2111-3360\log2)\alpha_{4}^{2}+270\alpha_{8}]$\tabularnewline
\hline 
$s_{10}$ & $\frac{4}{1575}(2999-5040\log2)\alpha_{4}\alpha_{6}+\frac{2}{9}\alpha_{10}$ & $\frac{13}{4725}[(2999-5040\log2)\alpha_{4}\alpha_{6}+175\alpha_{10}]$\tabularnewline
\hline 
\end{tabular}

\caption{Results for the expansion coefficients of the NG action for the circular
loop in powers of the radius $a$. }
\label{tab-comp-1}
\end{table}
These two approaches where considered in \cite{Drukker:1999zq} for
the case of the supersymmetric conformal theory. They essentially
differ in the regularization employed. In \cite{Drukker:1999zq} it
is shown that in the AdS case for smooth surfaces both regularizations
lead to the same results, except in what respects to zig-zag symmetry.
The HJ-regularization respects this symmetry but the $\epsilon$-regularization
does not. Bellow, these regularizations are compared for the non-AdS
case. 

Clearly the results appearing in the two colums in table \ref{tab-comp-1}
are different. Bellow, it is shown that the two computations would
reduce to a single one if an interchange of limits and integration
would be valid, which is not the case. 

The results for the $HJ$-regularization can be computed either by
the HJ approach or solving the differential equation and replacing
the solution in the NG action. Both methods lead to the same results.
In the second approach there appear terms in the integrand that go
to zero when $z_{1}\to0$ but survive after integration. These terms
are responsable for the discrepancy with the $\epsilon$-regularization
which, as mentioned before, is equivalent to taking the limit $z_{1}\to0$
in the integrand before performing the integral. In appendix B a concrete
example is considered which shows how these terms arise for  the case
of the coefficient $s_{2}$.

The main difference between both approaches is that, in the HJ-regularization,
boundary conditions for the minimal surface are taken at its border,
i.e. where the base loop lies. In the $\epsilon$-regularization boundary
conditions are taken at $z=0$ , which is not the location of the
calculated area border. This implies that in this last case, the calculted
area does not correspond to the area of a minimal surface, whose border
lies at $z=0$, but to a fraction of it. In the limit $\epsilon\to0$
the difference would vanish, however the divergence of the metric
for $z\to0$, gives a non vanishing contribution, which accounts for
the difference between both results. In this respect it is worth noting,
that such difference is not seen in the AdS case, simply because in
that case conformal invariance requires the vanishing of the condensates.

\section{Concluding remarks}

In this work the HJ approach has been employed for the calculation
of minimal areas on asymptotically AdS spaces. These calculations
are relevant, from the holographic point of view, in obaining expectation
values of Wilson loops in the gauge theory living at the border of
these spaces. In this respect it is worth noting that,
\begin{itemize}
\item This approach directly calculates the minimal area without need to
solve the equations of motion and replace the solution in the NG action.
This makes the calculation more direct and in practice much simpler.
\item In this approach variations of the on-shell classical action under
changes in its boundary conditions are studied. The location of the
loop countour is one of these conditions. Therefore the HJ-approach
also leads to a natural regularization, which consists in moving the
location of the countour out of the border.
\end{itemize}

Regarding the issue of regularization schemes it was shown that different
schemes lead to different results. If one requires zig-zag symmetry
to be respected then, as shown in \cite{Drukker:1999zq}, the HJ scheme
should be choosen. In this respect it is important to note that the
HJ-scheme for any value of the regularization parameter $z_{1}$,
computes the area of a minimal surface. This is not the case for the
$\epsilon$-scheme.

\section*{Appendix A: The first terms in the expansion in powers of the radius.}

\begin{eqnarray*}
C^{(2)}(x) & = & -\frac{\beta_{2}}{x^{2}}\,\textrm{,}\\
s_{2}(w_{1}) & = & -w_{1}\,\sqrt{1+w_{1}^{2}}\,\frac{\beta_{2}}{2}+\frac{\beta_{2}}{2}\,\left(1+w_{1}^{2}\right).
\end{eqnarray*}

\begin{eqnarray*}
C^{(4)}(x) & = & \left(1+x^{2}\right)\,s'_{2}(x)^{2}-4\,x\,s'_{2}(x)\,s_{2}(x)+4\,s_{2}(x)^{2}-\beta_{4}\,\textrm{,}\\
s_{4}(w_{1}) & = & \frac{1}{24}\left\{ 4\beta_{4}+4\beta_{4}\left(2+w_{1}^{2}\right)w^{2}-4\beta_{4}\,w_{1}^{3}\,\sqrt{w_{1}^{2}+1}\right.\\
 &  & +3\beta_{2}^{2}\left[w_{1}^{2}\left(9w_{1}^{2}-9w_{1}\sqrt{w_{1}^{2}+1}+14\right)-8\sqrt{w_{1}^{2}+1}+5w_{1}\right]\\
 &  & \left.+12\beta_{2}^{2}\left(1+w_{1}^{2}\right)^{2}\left[2\,\arcsin(w_{1})-\log\left(1+w_{1}^{2}\right)-\log\left(4\right)\right]\right\} \,\textrm{.}
\end{eqnarray*}

\begin{eqnarray*}
C^{(6)}(x) & = & 2\left(1+x^{2}\right)\,s'_{2}(x)\,s'_{4}(x)-8x\,s_{4}(x)\,s'_{2}(x)-4x\,s{}_{2}(x)\,s'_{4}(x)+16\,s{}_{2}(x)\,s{}_{4}(x)-\beta_{6}\,x^{2}\,\textrm{,}\\
s_{6}(w_{1}) & = & -\frac{1}{48}\left(1+w_{1}^{2}\right)^{3}\left\{ -\frac{24}{5}\beta_{6}-24\beta_{2}^{3}\,\log^{2}(4)+3\beta_{2}^{3}\left[36\,\log(4)-73\right]+\frac{4}{3}\beta_{2}\beta_{4}\left[24\log(4)-59\right]\right.\\
 &  & -\frac{8\left(3\beta_{2}^{2}+\beta_{4}\right)\beta_{2}}{\left(1+w_{1}^{2}\right)^{2}}+\frac{204\beta_{2}^{3}+48\beta_{2}\beta_{4}}{1+w_{1}^{2}}-24\beta_{2}^{3}\,\log^{2}\left(1+w_{1}^{2}\right)-96\beta_{2}^{3}\,\log(2)\,\log\left(1+w_{1}^{2}\right)\\
 &  & +4\left(27\beta_{2}^{3}+8\beta_{2}\beta_{4}\right)\,\log\left(1+w_{1}^{2}\right)+\frac{192\beta_{2}^{3}\,w_{1}\,\arcsin(w_{1})}{\sqrt{1+w_{1}^{2}}}+\frac{96\beta_{2}^{3}\,\arcsin(w_{1})}{1+w_{1}^{2}}\\
 &  & -\frac{48\beta_{2}^{3}\,\log\left(4+4w_{1}^{2}\right)\left(1+2w_{1}\,\sqrt{1+w_{1}^{2}}-2\left(1+w_{1}^{2}\right)\arcsin(w_{1})\right)}{w^{2}+1}\\
 &  & -96\beta_{2}^{3}\,\arcsin^{2}(w_{1})-216\beta_{2}^{3}\,\arcsin(w_{1})-64\beta_{2}\beta_{4}\,\arcsin(w_{1})\\
 &  & \left.+w_{1}\,\frac{72\beta_{6}\,w_{1}^{4}+45\beta_{2}^{3}\left(73w_{1}^{4}+112w_{1}^{2}+40\right)+20\beta_{2}\beta_{4}\left(59w_{1}^{4}+104w_{1}^{2}+48\right)}{15\left(1+w_{1}^{2}\right)^{5/2}}\right\} \,\textrm{.}
\end{eqnarray*}

\section*{Appendix B: The difference between the $z_{1}$-regularization and
the $\epsilon$-regularization}

\noindent In terms of the variables,
\[
t=\sqrt{1+\left(\frac{z_{1}}{a}\right)^{2}-\left(\frac{r}{a}\right)^{2}}\;\;,\psi(t)=\left(\frac{z}{a}\right)^{2}
\]
the NG action is,
\begin{equation}
S_{NG}=\frac{L^{2}}{\alpha'}\int_{w_{1}}^{\sqrt{1+w_{1}^{2}}}\frac{e^{2(a^{2}\alpha_{2}\psi+a^{4}\alpha_{4}\psi^{2})}t\sqrt{4+\frac{\left(1+w_{1}^{2}-t^{2}\right)\psi'(t)^{2}}{t^{2}\psi\left(t\right)}}}{2\psi\left(t\right)}dt\;.\label{eq:NG en funcion de t-1}
\end{equation}
the solution to the equation of motion with the boundary conditions,
\[
\psi(w_{1})=w_{1}^{2}\;\equiv(z(a)=z_{1})\;,\;\psi'(\sqrt{1+w_{1}^{2}})=0\;,\equiv(z'(0)=0)\text{ }
\]

\noindent up to order $a^{2}$ is given by,

\begin{eqnarray*}
\psi(t) & = & t^{2}-4a^{2}\alpha_{2}\left(w_{1}^{2}+1\right)\left(\left(w_{1}^{2}+1\right)\log\left(-t^{2}+w_{1}^{2}+1\right)+\right.\\
 &  & \left.+(w_{1}-t)\left(-t+2\sqrt{w_{1}^{2}+1}-w_{1}\right)\right.\\
 &  & \left.+2\left(w_{1}^{2}+1\right)\tanh^{-1}\left(\frac{t}{\sqrt{w_{1}^{2}+1}}\right)-2\left(w_{1}^{2}+1\right)\sinh^{-1}(w_{1})\right)
\end{eqnarray*}
replacing in (\ref{eq:NG en funcion de t-1}) gives the following
expression for the integrand up to order $a^{2}$,
\begin{eqnarray*}
I(w_{1},t) & = & \frac{\sqrt{1+w_{1}^{2}}}{t^{2}}+a^{2}\frac{2\alpha_{2}\sqrt{w_{1}^{2}+1}}{t^{4}}\left(2t^{4}+4t^{2}w_{1}^{2}-2t^{2}w_{1}\sqrt{w_{1}^{2}+1}\right.\\
 &  & \left.-t^{2}w_{1}^{2}\log\left(-t^{2}+w_{1}^{2}+1\right)-t^{2}\log\left(-t^{2}+w_{1}^{2}+1\right)+6w_{1}^{2}\log\left(-t^{2}+w_{1}^{2}+1\right)\right.\\
 &  & \left.+3\log\left(-t^{2}+w_{1}^{2}+1\right)+2\left(w_{1}^{2}+1\right)\left(t^{2}-3\left(w_{1}^{2}+1\right)\right)\sinh^{-1}(w_{1})\right.\\
 &  & \left.-2\left(w_{1}^{2}+1\right)\left(t^{2}-3w_{1}^{2}-3\right)\tanh^{-1}\left(\frac{t}{\sqrt{w_{1}^{2}+1}}\right)+3w_{1}^{4}\log\left(-t^{2}+w_{1}^{2}+1\right)\right.\\
 &  & \left.+3t^{2}-6t\,w_{1}^{2}\sqrt{w_{1}^{2}+1}-6t\sqrt{w_{1}^{2}+1}-3w_{1}^{4}-3w_{1}^{2}+6w_{1}\sqrt{w_{1}^{2}+1}+6w_{1}^{3}\sqrt{w_{1}^{2}+1}\right)
\end{eqnarray*}
the last term in this integrand is proportional to $w_{1}^{3}/t^{4}$,
which vanish when $w_{1}\to0$. However integrating and then taking
the limit, they lead to a non-vanishing result,
\[
\int_{w_{1}}^{\sqrt{1+w_{1}^{2}}}dt\;\frac{w_{1}^{3}}{t^{4}}=\left.-\frac{1}{3}\frac{w_{1}^{3}}{t^{3}}\right|_{w_{1}}^{\sqrt{1+w_{1}^{2}}}\stackrel{_{w_{1}\to0}}{=}\frac{1}{3}\,\textrm{.}
\]

\bibliographystyle{unsrt}
\bibliography{Bibliography}

\end{document}